\documentclass[12pt]{article}
\newcommand{\be}{\begin{equation}}
\newcommand{\ee}{\end{equation}}
\newcommand{\ba}{\begin{eqnarray}}
\newcommand{\ea}{\end{eqnarray}}

\begin{document}
\setlength{\baselineskip}{.7cm}
\renewcommand{\thefootnote}{\fnsymbol{footnote}}
\newcommand{\lp}{\left(}
\newcommand{\rp}{\right)}

\begin{center}
\centering{{\bf \Large Acoustic fluidization for earthquakes?\\}
\vskip 1cm
Didier Sornette $^{1,2}$ and Anne Sornette $^2$\\
$^1$ Institute of Geophysics and Planetary Physics\\ and
Department of Earth and Space Sciences\\
UCLA, Box 951567, Los Angeles, CA 90095-1567\\
$^2$ Laboratoire de Physique de la Mati\`ere Condens\'ee, CNRS UMR6622\\
Universite de Nice-Sophia Antipolis, 06108 NICE Cedex 2, France}
\end{center}

\vskip 3cm

Abstract\,: Melosh [1996] has suggested that acoustic fluidization could provide an 
alternative to theories that are invoked as explanations
for why some faults appear to be weak. We show that there is a subtle but profound
inconsistency in the theory that unfortunately invalidates the 
results. We propose possible remedies but must acknowledge that the relevance
of acoustic fluidization remains an open question.

\pagebreak

\section{Introduction}

In the standard rebound theory of earthquakes,
deformation elastic energy is progressively stored in the crust and is
suddenly released in an earthquake when a threshold is reached.
The Ruina-Dieterich friction laws [{\it Dieterich}, 1972; 1978; {\it Ruina}, 1983] 
constitute the basic ingredient used to
describe the interaction between the two sides of the sliding fault. 
Friction coefficients based on laboratory experiments [{\it Byerlee}, 1977;
{\it Scholz}, 1998] fail to account for 
modern observations of strain [{\it Jackson et al.}, 1997], stress 
[{\it Zoback et al.}, 1987; {\it Zoback}, 1992a; 1992b] and heat flows
[{\it Henyey and Wasserburg}, 1971; {\it Lachenbruch and Sass}, 1980; 1988; 1992;
 {\it Lachenbruch et al.}, 1995] (see Sornette [1999] for a synthesis). 
 
Resolutions of these
paradoxes usually call for additional mechanisms, involving fluids 
[{\it National Research Council}, 1990], crack opening modes of 
slip [{\it Brune et al.}, 1993], dynamical collision effects [{\it Lomnitz}, 
1991; {\it Pisarenko and Mora}, 1994], frictional properties of a granular gouge 
model under large slip [{\it Scott}, 1996], space filling bearings with compatible 
kinematic rotations [{\it Herrmann et al.}, 1990], hierarchical scaling [{\it 
Schmittbuhl et al.}, 1996], etc.

Melosh [1996] has recently suggested that the mechanism of 
``acoustic fluidization'' could provide an 
alternative to theories that invoke pressurized fluids as an explanation
for why some faults appear to be weak. 
Fluidization usually refers to the experimental observation
that granular material in the presence of an interstitial fluid can liquidify when
shaken sufficiently strongly [{\it Russo et al.}, 1995].
The liquifaction is due to the fact that granular media are first compressive for
small deformation leading to an increase of the interstitial fluid pressure. This
increase in turn decreases the friction between the grains that can eventually become
free to shear. For example, liquifaction of sediments by resonant amplified
seismic waves have been proposed to be in part responsible for the damage and collapse
of certain buildings during the Michoacan earthquake, 1985 [{\it Lomnitz}, 1987]
and for the damage in the Marina district of San Francisco during the Loma Prieta earthquake
[{\it Bardet}, 1990]. 

In the acoustic fluidization mechanism [{\it Melosh}, 1979; 1996], 
no interstitial fluid is invoked. A fraction $e$ of the earthquake energy is released
as high-frequency acoustic waves that scatter off and shake the granular gouge
leading to the build-up of a local acoustic pressure. According to 
[{\it Melosh}, 1996], when this pressure 
becomes of the order of the overburden lithostatic
pressure $\rho g h$, 
the granular gouge becomes essentially free to slip without much residual friction.

The purpose of this note is to show that there is a problem with this mechanism because it
predicts a slip velocity during an earthquake more than two orders of magnitudes
smaller than the typical meters per second for observed earthquakes, in contradiction
to the result of Melosh [1996]. The problem
stems from a confusion in the definition of dissipation and scattering lengths.
We then suggest possible modifications of Melosh's theory that could
resolve this problem and which lead to a richer theory.

\section{Summary of Melosh's theory and useful background}

\subsection{Acoustic wave energy transport}

The first ingredient is the generation and transport of high-frequency acoustic waves in 
the core of the fault. Melosh [1996] uses the standard diffusion equation (his equation 2)
for the  elastic transport of acoustic waves [{\it Ishimaru}, 1978; {\it Sornette}, 1989a-c] 
with a dissipation and a source term\,:
\be
{d E \over dt} = {\xi \over 4} \nabla^2 E - {c \over \lambda Q} E + 
e {{\dot \epsilon} \tau \over 2}~,
\label{hqhqjq}
\ee
where $E$ is the acoustic wave elastic energy density. The diffusion
coefficient $\xi/4$, where $\xi$ is called the scattering diffusivity by Melosh [1996],
can be expressed in terms of the 
elastic mean free path $l_e$ and of the transport 
acoustic wave velocity $c$ at scales below $l_e$. The velocity 
$c$ is of the order of the shear wave velocity 
[{\it Turner and Weaver}, 1996; {\it Van Albada et al.}, 1991; 
{\it Van Tiggelen and Lagendijk}, 1993]. We thus get 
[{\it Ishimaru}, 1978; {\it Sornette}, 1989a,c] \,:
\be
{\xi \over 4} \simeq {1 \over 3} c l_e ~.
\label{hhdqkkq}
\ee
We stress that the term ``elastic'' refers to the fact that $l_e$ is the characteristic
distance over which an acoustic wave propagates before being scattered in other
directions, {\it without} any loss of energy. 

The l.h.s. and first term of the r.h.s. of (\ref{hqhqjq}) give the diffusion equation
which describes the transport of an acoustic wave in a multiple-scattering medium.
The second term of the r.h.s. of (\ref{hqhqjq}) will be shown to describe the
presence of a genuine absorption, while the last source term corresponds to the 
conversion to acoustic waves of a fraction $e$ of the mechanical work
performed per unit time
by the fault motion with strain rate ${\dot \epsilon}$ and shear stress $\tau$.

\subsection{Diffusive transport}

The first two terms of (\ref{hqhqjq}) 
\be
d E / dt = (\xi / 4) \nabla^2 E
\ee
gives the
standard parabolic diffusion equation which is 
based on the following processes. Once generated from a source, 
an acoustic wave propagates roughly ballistically over a typical distance
of the order of the elastic (scattering) mean free path $l_e$. Over
this distance, the equation governing the acoustic wave propagation
is the hyperbolic wave equation
\be
{\partial^2 A \over \partial t^2} = c^2~\nabla^2 A~,  \label{jfjjjds}
\ee
where the wave amplitude $A$ is related to $E$ by $E=|A|^2$ and the wave velocity
$c$ may depend locally on position to reflect the heterogeneity of the medium.
Due to this heterogeneity, the wave is scattered off its initial propagation path
along the direction $x$
and its intensity in this direction $x$ decays as $\exp[-x/l_e]$. 
This exponential decay of the intensity does not correspond to a genuine absorption but
rather reflects the loss of acoustic energy along the
direction $x$ to all possible scattered waves in all other directions. Mathematically,
the exponential decay $\exp[-x/l_e]$ can be derived from (\ref{jfjjjds})
using standard scattering theory [{\it Ishimaru}, 1978]. The
conservation of acoustic energy is ensured by the fact that the sum of wave
intensity over all directions of propagation remains constant.

Beyond the distance $l_e$, the nature of the transport of the wave intensity crosses over
from ballistic (i.e. straight propagation) to diffusive, corresponding to the 
picture where the acoustic wave can be viewed as a superposition of random walks with 
typical step length equal to $l_e$. 
This means that the equation for the wave propagation changes from 
the hyperbolic wave equation for the wave amplitude to the 
parabolic diffusion equation for the wave intensity 
given by the first two terms of (\ref{hqhqjq}). 

One can quantify this by the following example. Consider an acoustic wave of energy $E_0$
impinging on a slab of thickness $L$ made of heterogeneities that scatter off the
acoustic wave, and whose scattering strength is quantified by the elastic
mean free path $l_e$. Anderson [1985] and
Sornette [1989c,d] have revisited this diffusion equation
to get the transmission coefficient in this example, i.e. the acoustic energy which is 
transmitted to the other size of the slab, as a function of its thickness $L$. The
result is\,:
\be
E(L) \simeq   E_0~ {l_e \over L}~.  
\label{gqgqhjqjj}	
\ee
Note that the decay follows the algebraic $1/L$ law rather than an exponential law.
Furthermore, the acoustic intensity profile within the slab is linear 
and not exponentially decreasing\,:
\be
E(z) \simeq  E_0 {L+l_e/3 - x \over L}~,~~~~~~{\rm for}~~ l_e < x < L-l_e~.
\label{fqgqgqgp}
\ee
These results (\ref{gqgqhjqjj},\ref{fqgqgqgp}) highlight that the diffusive
transport of the acoustic energy due to multiple scattering event is very
different from the exponential attenuation that a genuine absorption would produce.

\subsection{Absorption}

The third term $-(c/\lambda Q) E$ of (\ref{hqhqjq}) quantifies genuine absorption processes.
The parameter $\lambda$ is the acoustic wavelength and $Q$ is the quality factor.
To see that this term reflects absorption, we consider (\ref{hqhqjq}) in absence of the 
spatial derivative $\nabla^2 E$ and of the last source term\,:
\be
{d E \over dt} = - {c \over \lambda Q} E ~.
\label{hqhqqqjq}
\ee
Its solution is 
\be
E(t) = E_0 ~\exp\left(-{c \over \lambda Q}~t\right)~,
\ee
which is very different from the energy decay (\ref{gqgqhjqjj}) solely due to 
diffusion. It is thus clear that the term $ -(c/\lambda Q) E$ is not
coming from elastic scattering but solely from genuine absorption, 
i.e. conversion of acoustic energy into thermal energy.

The usual definition of the quality factor $Q$ is [{\it Knopoff}, 1964]
\be
Q \equiv 2\pi {l_a \over \lambda}~,
\label{gqggqhqj}
\ee
where $l_a$ is the absorption length defined by the exponential decay $\exp 
\left( -x/l_a \right)$ of a {\it ballistically} propagating wave in an absorbing medium.

Melosh [1996] introduces a characteristic length $l_*$, which he calls (misleadingly)
the ``scattering length'', defined by
\be
l_* \equiv \sqrt{\xi \lambda Q \over 4 c}~.
\label{hqqlmmq}
\ee
Using (\ref{hhdqkkq}) and (\ref{gqggqhqj}), we get
\be
l_* \simeq \sqrt{2\pi \over 3}~~\sqrt{l_e l_a}~.
\label{qffqqdd}
\ee
Calling $l_*$, a ``scattering length'', is misleading because $l_*$ is in reality
the effective absorption length in the diffusive medium. To see this, we use
the standard diffusion relation 
\be
l_*^2 \approx 6 {\xi \over 4} \tau_a
\ee
linking the radius of gyration $l_*$ covered by a diffusing process over a time $\tau_a=l_a/c$
equal to the time
needed for the wave to cover the real distance $l_a$, along its convoluted multi-scattered
path. The prefactor $6$ holds for diffusion in a three dimensional space.

Using (\ref{hhdqkkq}), we get $l_* \approx \sqrt{2l_e l_a}$, which recovers the 
(\ref{qffqqdd}) up to a numerical factor of order unity. 
The expression (\ref{qffqqdd}) can be derived by several other 
methods [{\it Sornette}, 1989d]. What is important is that $l_*$ scales as the geometrical
mean of $l_e$ and $l_a$, which comes from the random walk nature of the diffusive process.
Physically, in the diffusive regime,
the acoustic wave energy is absorbed over
the characteristic length $l_*$, which stems from the fact that, to cross the distance
$l_*$, the wave follows random walk paths of length $l_a \sim l_*^2/l_e$.
This reflects that attenuation of a wave in a scattering medium
is a function of both absorption of energy and scattering.

\subsection{Feedback of the acoustic vibrations on the slip rate}

The interesting idea of Melosh [1996] is that the high-frequency vibrations
may shake the fault and unlock it, leading to an easier sliding motion. For this,
he proposes the following effective friction equation, relating the strain rate
${\dot \epsilon}$, the shear stress $\tau$ and the normalized 
acoustic wave energy $\Psi = E~\rho c^2/(\rho g h)^2$\,:
\be
{\dot \epsilon} = {\tau \over \rho \lambda c}~ \biggl[ {1 - {\rm erf}({1 \over 2\sqrt{\Psi}})
\over 1 + {\rm erf}({1 \over 2\sqrt{\Psi}})} \biggl]~.
\label{aaqsqqqwq}
\ee
The main physical phenomenon taken into account in this equation is that, due to the
acoustic shaking, the effective viscous friction 
$\tau/{\dot \epsilon}$ is a decreasing function of the acoustic wave energy.
This mechanism is related to the velocity weakening mechanism induced by
collision between asperities that lead to a transfer of momentum from the direction
parallel to the fault to the direction transverse to it [{\it Lomnitz-Adler}, 1991;
{\it Maveyraud et al.}, 1998].

Putting (\ref{aaqsqqqwq}) in (\ref{hqhqjq}) and looking for stationary
modes gives the non-linear ordinary differential equation (\ref{aaqswq})
(for the case $\eta = 1$, see below), 
whose analysis leads to the 
prediction of two rupture modes [{\it Melosh}, 1996].

\section{Problem with Melosh's theory}

In order to obtain realistic values, there are some constraints that the 
model parameters must satisfy. The key parameter is the ``regeneration'' parameter
\be
r \Sigma^2 = {e Q \over 2}~\left({\tau \over\rho g h}\right)^2~,
\ee
where $\Sigma = \tau/\rho g h$ is the normalized shear stress.
Melosh [1996] finds
reasonable solutions only for $r \Sigma^2 > 2.8$.
For a typical fraction $e \approx 0.1$ of conversion to acoustic waves of the mechanical work
performed per unit time by the fault motion and for a ratio $\tau/\rho g h$ as low as 
$0.1$ as suggested from observations on the San Andreas fault, this value
$r \Sigma^2 > 2.8$ corresponds to $Q > 5600$. This 
estimation may vary by an order of magnitude with the conversion factor and the
relative shear stress. However, the message is that the quality factor
$Q$ measuring the attenuation of 
the acoustic waves must be high, in the range of $10^3$ for
the acoustic waves to be self-sustained during the earthquake slip motion. 
This is the first condition.

On the other hand, Melosh's theory predict the slip velocity 
\be
{\dot u} \approx 1.4~{\tau \over \rho c}~ {l_* \over \lambda}
\label{gqkqkkll}
\ee
during a typical earthquake. Using a shear stress $\tau \approx 10~MPa$, 
a density $\rho = 3000~kg/m^3$
and $c=4~km/s$ gives ${\dot u} \approx 1.2~(l_* /\lambda)~m/s$. Thus, a realistic
slip velocity ${\dot u} \approx 1~m/s$ requires that
\be
l_* \approx \lambda~.
\ee
Together with (\ref{qffqqdd}) and (\ref{gqggqhqj}), this leads to 
\be
Q \approx 3~{\lambda \over l_e}~.
\label{hjlllm}
\ee
This last expression (\ref{hjlllm}) is totally incompatible with the above condition
$Q \geq 10^3$, as this would lead to $l_e \approx \lambda/100$ or smaller. This last
condition is a physical impossibility\,: the elastic scattering length is always much
larger than or at the extreme limit of the same order as the wavelength. The
physical intuition is that a wave is defined over a length scale of the order of the
wavelength (otherwise, there are no spatial oscillations) and the scattering process
needs at least this scale to operate. The limit $l_e \simeq \lambda$
is attained only under exceptional circumstances leading to a novel phenomenon, called
Anderson localization, in which the acoustic wave do not propagate anymore but oscillate
locally. Extraordinary efficient scatterers are needed to reach this regime 
[{\it Sornette}, 1989c]. It is thus clear that the condition $l_e \approx \lambda/100$ 
is utterly unphysical.

If in constrast, we put $Q \approx 10^3$ in (\ref{hqqlmmq}), we get
$l_* \approx 160~\lambda$, which from (\ref{gqkqkkll})
 leads to a maximum slip velocity ${\dot u} \approx
7.5~mm/s$, using the numerical example of Melosh [1996]. This slow sliding velocity
is unrealistic for earthquakes.

\section{Possible remedies}

A first remedy is to relax the condition used by Melosh that
the acoustic pressure needs
to reach the overburden pressure in order to significantly affect the
fault friction. We propose that only a small
fraction $\eta$ of it is enough to liquidify the fault. 
 
 Indeed, it is well-established experimentally 
[{\it Biarez and Hicher}, 1994] that the elastic modulii 
of granular media under large cyclic
deformations are much lower than their static
values. This effect occurs only for sufficiently large amplitudes of the cyclic
deformation, typically for strains $\epsilon_a$ above $10^{-4}$. 
At $\epsilon_a = 10^{-3}$, the elastic modulii
are halved and at $\epsilon_a = 10^{-2}$, the elastic modulii
are more than five times smaller than their static values. As a consequence, the
strength of the granular medium is decreased in proportion. 
Melosh (private communication) also finds in laboratory experiments that
a large decrease in elastic modulus is required to fit the measured flow rate
of acoustically fluidized debris. This is consistent with flow in granular
material in a completely dilatent state and agrees with measurements reported
in the literature for the elastic modulus of highly strained granular materials.

In absence of cohesion forces, the strength of a granular material is solely
due to the effect of gravity weight that put grains in contact together
and the resistance to shear is
governed by Coulomb's law according to which the shear stress at the threshold
for sliding
is proportional to the normal stress. The coefficient of proportionality is the
friction coefficient. As we have discussed above,
in presence of acoustic fluidization, the resistance
to shear deformation is decreased as a consequence of the reduction
of the effective elastic modulus.
Correlatively, the threshold for sliding is also decreased (strength
is often proportional to elastic modulus in brittle material). In order to capture
these phenomena, we propose, what is maybe the simplest approximation, that
the criterion for unlocking the fault is changed from Melosh's criterion
to the condition that the acoustic pressure needs only reach a fraction $\eta$
of the overburden pressure. Let us stress that the essence of 
our argument is that the acoustic energy is fed by the moving fault and thus
the acoutic particle velocity adjust to the changing elastic modulus so that 
the overall acoustic energy is ``externally'' controlled by the rate of 
global elastic dissipation. The acoustic fluidization thus controls the
sliding threshold rather than solely the acoustic particle velocity.

We need 
to estimate the strain created by the acoustic field. The acoustic
pressure is related to the acoustic particle velocity $v$ by 
\be
p = \rho c v~.
\ee
Assuming 
\be
p = \eta \rho g h~,
\ee
 this yields 
 \be
 v = \eta g h/c \approx 12~m/s
 \ee
  for
$p \approx 200~MPa$, a density  $\rho = 3~10^3~kg/m^3$, a velocity 
$c=4000~m/s$ and $\eta = 0.1$. 
This corresponds to an acoustic wave displacement $u_a = 
v/2 \pi f \approx 2~10^{-3}~m$ at a frequency $f \sim 10^3~Hz$.
 The corresponding strain $u_a/w$ is $\sim 2~10^{-3}$
for a gouge width $w$ of the order of one meter [{\it Melosh}, 1996] over which the
intense shaking occurs. These estimations suggest that Melosh's criterion
that the acoustic stress fluctuations must approach the overburden stress on the fault
for acoustic fluidization to occur
is too drastic and smaller shaking can reduce significantly the fault friction.

Persuing this reasoning, we see that the fundamental equation (6) in [{\it Melosh}, 1996]
is changed into
\be
{d^2 \Psi \over d\zeta^2} = \Psi - r\Sigma^2 \biggl[ {1 - erf({\eta \over 2\sqrt{\Psi}})
\over 1 + erf({\eta \over 2\sqrt{\Psi}})} \biggl]~,
\label{aaqswq}
\ee
where $\eta = 1$ recovers the case treated by Melosh.
$\Psi$ is the normalized acoustic energy, $\zeta = z/l_*$, $z$ is the coordinate
perpendicular to the fault, the regeneration parameter is $r = eQ/2$ where $e$ is the
acoustic energy conversion efficiency, $\Sigma = \tau/ \rho g h$ and erf$(x)$ is the
error function. 
We see that a factor $\eta<1$ implies a more effective
generation of acoustic waves because the second source term of the r.h.s. is larger. But,
since the bracket term saturates to one for large energies and/or small $\eta$,
this does not lead to a significantly 
larger slip velocity than found above. This remains a 
problem of the theory. 

This problem might be alleviated by
treating $e$ self-consistently 
as a decreasing function of the friction coefficient, and thus of the acoustic energy density.
The problem then becomes even more non-linear because it reflects 
in addition the dependence of the acoustic radiation efficiency of the granular gouge
on the amplitude of the acoustic vibrations. In addition, the elastic modulus is
also really nonlinear and it is only the tangent modulus that decreases close to the 
yield in the dilatent region, which suggests that the above linear formalism is not
adequate and should be modified.

Further 
improvement could also take into account that the stochastic acoustic energy 
may deviate from a Rayleigh distribution [{\it Ishimaru}, 1978; {\it Mirlin et al.}, 1998]
when the medium is strong heterogeneous. This
modifies the functional form of the term in bracket in eq.(\ref{aaqswq}) and
thus all numerical estimations.

We ackowledge stimulating discussions and correspondence with H.J. Melosh.

\pagebreak

{\bf References}\,:
\vskip 0.3cm

Anderson,~P.W., (1985). The question of classical localization. A theory of white paint?
{\it Phil. Mag. B},~{\bf 52}, 505-509.

Bardet,~J.P., (1990). Damage at a distance, {\it Nature},~{\bf 346},~799-799.

Biarez,~J. and Hicher,~P.-Y., (1994). 
Elementary mechanics of soil behavior\,: saturated
remoulded soils, Rotterdam ; Brookfield, VT : A.A. Balkema.

Brune, J.N., S. Brown and P.A. Johnson, (1993).
Rupture mechanism and interface separation in foam rubber models of
earthquakes -  A possible solution to the heat flow paradox and the paradox of large
overthrusts, {\it Tectonophysics},~{\bf 218},59-67.

Byerlee,~J., (1977). Friction of rocks, In Experimental studies of rock friction with
application to earthquake prediction, ed. J.F. Evernden, U.S. Geological Survey,
Menlo Park, Ca, 55-77.

Dieterich,~J.H., (1972). Time-dependent friction in rocks,
{\it J. Geophys. Res.},~{\bf 77}, 3690-3697.

Dieterich,~J.H., (1978). Time-dependent friction and the mechanics of stick-slip,
{\it Pure and Applied Geophysics},~{\bf 116}, 790-806.

Henyey,~T.L., and Wasserburg,~G.J., (1971). Heat flow near major strike-slip
faults in California, {\it J. Geophys. Res.},~{\bf 76}, 7924-7946.

Herrmann,~H.J., G. Mantica and D. Bessis, (1990). Space-filling bearings,
{\it Phys.Rev.Lett.}, {\bf 65},~3223-3226.

Ishimaru,~A., (1978). Wave propagation and scattering in random media
Academic Press, New York.

Jackson,~D.D., Shen,~Z.K., Potter,~D., Ge,~X.B. and others, (1997).
Southern California deformation, {\it Science},~{\bf 277},~1621-1622.
 
Knopoff,~K., (1964). Q, {\it Reviews of Geophysics},~{\bf 2}, 625-660.

Lachenbruch,~A.H., and Sass,~J.H., (1980). Heat flow and energetics of the San
Andreas fault zone, {\it J. Geophys. Res.},~{\bf 85}, 6185-6222.

Lachenbruch,~A.H., and Sass,~J.H., (1988). The stress heat-flow paradox and thermal results 
from Cajon Pass, {\it Geophys. Res. Lett.},~{\bf 15}, 981-984.

Lachenbruch,~A.H., and Sass,~J.H., (1992). Heat flow from Cajon Pass, fault strength
and tectonic implications, {\it J. Geophys. Res.},~{\bf 97}, 4995-5015.

Lachenbruch,~A.H., Sass,~J.H., Clow,~G.D. and Weldon,~R., (1995). Heat flow at
Cajon Pass, California, revisited, {\it J. Geophys. Res.},~{\bf 100}, 2005-2012.

Lomnitz,~C., (1987). Nonlinear behavior of clays in the great Mexico earthquake,
{\it C. R. Acad. Sci. Paris II},~{\bf 305},~1239-1242. 

Lomnitz-Adler,~J., (1991). Model for steady state friction,
{\it J. Geophys. Res.},~{\bf 96},~6121-6131.

Maveyraud,~C., W. Benz, G. Ouillon, A. Sornette and D. Sornette, (1998).
Solid friction at high sliding velocities\,: an explicit 3D dynamical SPH approach,
J. Geophys. Res. 104 (B12), December 10 (1999)

Melosh,~H.J., (1979). Acoustic fluidization\,: a new geological process?
{\it J. Geophys. Res.},~{\bf 84},~7513-7520.

Melosh,~H.J., (1996).
Dynamical weakening of faults by acoustic fluidization, {\it Nature},~{\bf 379},~601-606.

Mirlin,~A.D., Pnini,~R., and Shapiro,~B., (1998). 
Intensity distribution for waves in disordered media: Deviations from
 Rayleigh statistics, {\it Phys. Rev.~E},~{\bf 57},~R6285-R6288.

National Research Council (1990). The role of fluids in crustal processes, Studies in
geophysics, Geophysics study committed, Commission on Geosciences,
Environment and
Ressources, National Academic Press, Washington D.C..

Pisarenko,~D., and P. Mora, (1994). Velocity weakening in a dynamical model of
friction, {\it Pure and Applied Geophysics},~{\bf 142},~447-466.

Ruina,~A., (1983). Slip instability and state variable friction laws,
{\it J. Geophys. Res.},~{\bf 88},~10359-10370.

Russo,~P., Chirone,~R., Massimilla,~L., and Russo,~S., (1995). The influence
of the frequency of acoustic waves on sound-assisted fluidization of beds
of fine particles, {\it Powder Technology},~{\bf 82}, 219-230.

Schmittbuhl, J., J.-P. Vilotte and S. Roux, (1996).
Velocity weakening friction : A renormalization approach,
{\it J. Geophys. Res.},~{\bf 101},~13911-13917.

Scholz,~C.H., (1998). Earthquakes and friction laws, {\it Nature},~{\bf 391},~37-42.

Scott~D., (1996). Seismicity and stress rotation in a granular model of
the brittle crust, {\it Nature},~{\bf 381}, N6583,~592-595.

Sornette,~D., (1989a). Acoustic waves in random media\,: I Weak disorder regime,  
{\it Acustica},~{\bf 67},~199-215.

Sornette,~D., (1989b). Acoustic waves in random media\,: 
II Coherent effects and strong disorder regime, {\it Acustica},~{\bf 67},~251-265.

Sornette,~D., (1989c). Acoustic waves in random media: III
Experimental situations,  {\it Acustica},~ {\bf 68}, 15-25 .

Sornette,~D., (1989d). Anderson localization and wave absorption, 
{\it J.Stat.Phys.},~{\bf 56}, 669-680.

Sornette,~D., (1999). Earthquakes: from chemical alteration to mechanical rupture, 
{\it Physics Reports},~{\bf 313}, 238-292.

Turner,~J.A. and Weaver,~R.L., (1996). Diffusive energy propagation
on heterogeneous plates -- Structural acoustics radiative transfer theory,
{\it J. Acoust. Soc. Am.},~{\bf 100},~3686-3695.

Van Albada,~M.P., Van Tiggelen,~B.A., Lagendijk,~A., and 
Tip,~A., (1991). Speed of propagation of classical waves in 
strongly scattering media, {\it Phys. Rev. Lett.},~{\bf 66},~3132-3135.

Van Tiggelen,~B.A. and Lagendijk,~A., (1993). Rigorous treatment of
the speed of diffusing classical waves, {\it Europhys. Lett.},~{\bf 23},~311-316.

Zoback,~M.L., (1992a). 1st-order and 2nd-order patterns of stress in the lithosphere - The
world stress map project, {\it J. Geophys. Res.},~{\bf 97},~11703-11728.

Zoback,~M.L., (1992b). Stress field constraints on intraplate seismicity in Eastern
North-America, {\it J. Geophys. Res.},~{\bf 97},~11761-11782.

Zoback,~M.L., Zoback,~V., Mount,~J., Eaton,~J., Healy,~J., et al., (1987). 
New evidence on the state of stress of the San Andreas fault zone, 
{\it Science},~{\bf 238},~1105-1111.

\end{document}